\documentstyle[preprint,floats,pra,aps,psfig]{revtex}

\newcommand{\doublespace}{
    \renewcommand{\baselinestretch}{1.6}\large\normalsize}

\newcommand{\bce}{\begin{center}}
\newcommand{\ece}{\end{center}}
\newcommand{\be}{\begin{equation}}
\newcommand{\ee}{\end{equation}}
\newcommand{\bea}{\vspace{0.25cm}\begin{eqnarray}}
\newcommand{\eea}{\end{eqnarray}}

\def\PRA{{Phys. Rev.} A }

\doublespace

\begin{document}

\title{{\LARGE {\bf New experimental test of Bell inequalities by the use of a
non-maximally entangled photon state }}}

\doublespace

\author{G.Brida, M.Genovese \footnote{ genovese@ien.it. Tel. 39 011 3919234, fax 39 011 3919259}, C.Novero}
\address{Istituto Elettrotecnico Nazionale Galileo Ferraris, Str. delle Cacce\\
91,I-10135 Torino, }
\author{E. Predazzi}
\address{Dip. Fisica Teorica Univ. Torino e INFN, via P. Giuria 1, I-10125 Torino }
\maketitle

\vskip 1cm
{\bf Abstract}
\vskip 0.5cm 

We report on the realisation of a new test of Bell inequalities using the
superposition of type I parametric down conversion produced in two different
non-linear crystals pumped by the same laser, but with different
polarisations.

We discuss the advantages and the possible developments of this
configuration.
\vskip 1.5cm

\vskip 2cm
PACS: 03.65.Bz

Keywords: Bell inequalities, non-locality, hidden variable theories 

\vspace{8mm}

Many experiments have already been devoted to a test of Bell inequalities 
\cite{Mandel,asp,franson,type1,type2}, leading to a substantial agreement with quantum
mechanics and disfavouring realistic local hidden variable theories.
However, due to the low total detection efficiency (the so-called "detection
loophole") no experiment has yet been able to exclude definitively realistic
local hidden variable theories, for it is necessary a further additional
hypothesis \cite{santos}, stating that the observed sample of particles
pairs is a faithful subsample of the whole. This problem is known as  detection or efficiency loophole. Considering the extreme relevance
of a conclusive elimination of local hidden variable theories, the research
for new experimental configurations able to overcome the detection loophole
is of a great interest.

A very important theoretical step in this direction has been achieved
recognising that for non maximally entangled pairs a total
efficiency larger than 0.67 \cite{eb} (in the limit of no background) is required to
obtain an efficiency-loophole free experiment, whilst for maximally
entangled pairs this limit rises to 0.81. However, it must be noticed that
for non-maximally entangled states the largest discrepancy between quantum
mechanics and local hidden variable theories is reduced: thus a compromise
between a lower total efficiency and a still sufficiently large value of
this difference will be necessary in a realisation of an experiment
addressed to overcome the detection loophole.

On the experimental side, in recent years the largest improvements have been
obtained by using parametric down conversion
(PDC) processes.

This technique has been employed, since its discovery, for generations of
"entangled" photon pairs, i.e. pairs of photons described by a common wave
function which by no means can be factored up into the product of two
distinct wave functions pertaining to separated photons. It is due
essentially to L. Mandel and collaborators \cite{Mandel} the idea of using such a state of the electromagnetic field to perform tests of quantum mechanics,
more specifically to test Bell's inequalities.

The generation of entangled states by parametric down conversion is
alternative to other techniques, such as the radiative decay of atomic
excited states, as it was in the celebrated experiment of A. Aspect et al. 
\cite{asp}, and overcomes some former limitations in the direction of propagation
of the conjugated photons. In fact the poor angular correlation of
atomic cascade photons is at the origin of a small total efficiency of this set
up, leading to a selection of a small subsample of the produced photons,
which makes impossible to eliminate the detection loophole in this
case. On the other hand, a very good angular correlation (better than 1 mrad)
of the two photons of the pair is obtained in the PDC process, giving the
possibility, at least in principle, to overcome the previous problem.

The entanglement on phase and momentum, which is directly produced in Type I
parametric down conversion can be used for a test of Bell inequalities using
two spatially separated interferometers \cite{franson}, as realised by \cite{type1}.
The use of beam splitters, however, strongly reduces the total quantum
efficiency.

In alternative, one can generate a polarisation entangled state \cite{ou}. It appears, however, that the creation of couples of photons entangled
from the point of view of polarisation, which is by far the most diffuse
case due to the easy experimental implementation, still suffers severe
limitations, as it was pointed out recently in the literature. The essence
of the problem is that in generating this state, half of the initial photon
flux is lost (in most of the used configurations), and one is, of necessity,
led to assume that the photon's population actually involved in the
experiment is a faithful sample of the original one, without eliminating the
efficiency loophole.

A scheme which allows no postselection of the photons  has been
realised recently, using Type II PDC, where a polarisation entangled state
is directly generated. This scheme has effectively permitted, at the
price of delicate compensations for having identical arrival time of the
ordinary and extraordinary photon, a much higher total efficiency than the
previous ones. It is, however, still far from the value $0.81$ for the total
efficiency. Also, some recent experiments studying equalities among
correlations functions rather than Bell inequalities \cite{dem} are 
far by solving these problems \cite{garuccio}. A large interest remains
therefore for new experiments increasing total quantum efficiency in order
to reduce and finally overcome the efficiency loophole.

Our experiment follows and develops \cite{napoli} an idea by Hardy \cite
{hardy} and contemplates the creation of a polarisation (non maximally-)
entangled states of the form

\begin{equation}
\vert \psi \rangle = {\frac{ \vert H \rangle \vert H \rangle + f \vert V
\rangle \vert V \rangle }{\sqrt {(1 + |f|^2)}}}  \label{Psi}
\end{equation}
(where $H$ and $V$ indicate horizontal and vertical polarisations
respectively) via the superposition of the spontaneous fluorescence emitted
by two non-linear crystals driven by the same pumping laser. The
crystals are put in cascade along the propagation direction of the pumping
laser and the superposition is obtained by using an appropriate optics.

We describe more in details the experimental set-up, with reference to the sketch of figure 1 below. The two crystals of LiIO3 (10x10x10 mm) are 250 mm apart, a distance smaller than the coherence length of the pumping laser. This guarantees indistinguishibility in the creation of a couple of photons in the first or in the second crystal. A couple of planoconvex lenses of 120 mm focal length centred in between focalises the spontaneous emission from the first crystal into the second one maintaining exactly, in principle, the angular spread. A hole of 4 mm diameter is drilled into the centre of the lenses in order to allow transmission of the pump radiation without absorption and, even more important, without adding stray-light, because of fluorescence and diffusion of the UV radiation.   This configuration, which performs as a so-called "optical condenser", was chosen among others as a compromise between minimisation of aberrations (mainly spherical and chromatic) and losses due to the number of optical components. The pumping beam at the exit of the first crystal is displaced from its input direction by birefringence: the small quartz plate (5 x5 x5 mm) in front of the first lens of the condensers compensates this displacement, so that the input conditions are prepared to be the same for the two crystals, apart from alignment errors. Finally, a half-wavelength plate immediately after the condenser rotates the polarisation of the Argon beam and excites in the second crystal a spontaneous emission cross-polarised with respect to the first one. The dimensions and positions of both plates are carefully chosen in order not to intersect the spontaneous emissions at 633 and 789 nm. With a phase matching angle of $51^o$ they are emitted at $3.5^o$ and $4^o$ respectively. 

The precise value of $f$, which determines how far from a maximally
entangled state ($f=1$) the produced state is, can be easily tuned by modifying
the pump intensity between the two crystals. This is a fundamental property,
which could permit to select the most appropriate state for the practical
realisation of a detection loophole free experiment.

The alignment (which is of fundamental importance for having a high
visibility and in principle constitutes the main problem of such a
configuration) has been profitably improved using  an optical amplifier
scheme, where a solid state laser is injected into the crystals together with
the pumping laser, an argon laser at 351 nm wavelength (see fig.1). Such a technique had already proved its
validity and it is of great help in the recognition of the directions of
propagation of correlated wavelengths photon and applied in our laboratory \cite{brida} for metrological studies.

The proposed scheme is well suited for leading to a further step toward a
conclusive experimental test of non-locality in quantum mechanics. The
analysis of the experiments realised up to now \cite{santos} shows in fact
that visibility of the wanted effect (essentially visibility of interference
fringes) and overall quantum efficiency of detection are the main parameters
in such experiments. One first advantage of the proposed configuration with
respect to most of the previous experimental set-ups is that all the
entangled pairs are selected (and not only $50 \%$ as with beams splitters).

For the moment, the results which we are going to discuss are still far from
a definitive solution of the detection loophole; nevertheless, being the first test of Bell inequalities using a non-maximally
entangled state, they
represents an important step in this direction. Furthermore, this
configuration permits to use any pair of correlated frequencies and not only
the degenerate ones. We have thus realised this test using for a first time
two different wave lengths (at $633$ and $789$ nm).

A somehow similar set-up has been realised recently in ref. \cite{Kwiat}.
The main difference between the two experiments is that in \cite{Kwiat} the
two crystals are very thin and in contact with orthogonal optical
axes: this permits a "partial" superposition of the two emissions with
opposite polarisation. This overlapping is mainly due to the finite dimension of the pump laser beam, which reflects into a finite width of each wavelength emission.
We think that the arrangement, described in the present paper is better, in that by fine tuning the crystals' and optics' positions a perfect superposition can be obtained for at least a couple of wavelength in PDC emission.
Moreover, the parametric amplifier trick allows to approach this ideal situation very closely. 
  Furthermore, in the experiment of ref. \cite{Kwiat}
the value of $f$ is in principle tunable by rotating the
polarisation of the pump laser, however this reduce the power of the
pump producing PDC already in the first crystal, while in our case the whole pump
power can always be used in the first crystal, tuning the PDC produced in
the second one.

As a first check of our apparatus, we have measured the interference
fringes, varying the setting of one of the polarisers, leaving the other
fixed. We have found a high visibility, $V=0.973 \pm 0.038$.

Our results are summarised by the value obtained for the Clauser-Horne sum,

\begin{equation}
CH=N(\theta _{1},\theta _{2})-N(\theta _{1},\theta _{2}^{\prime })+N(\theta
_{1}^{\prime },\theta _{2})+N(\theta _{1}^{\prime },\theta _{2}^{\prime
})-N(\theta _{1}^{\prime },\infty )-N(\infty ,\theta _{2})  \label{eq:CH}
\end{equation}
which is strictly negative for local realistic theory. In (\ref{eq:CH}), $N(\theta _{1},\theta _{2})$ is the number of coincidences between
channels 1 and 2 when the two polarisers are rotated to an angle $\theta _{1}$
and $\theta _{2}$ respectively ($\infty $ denotes the absence of selection
of polarisation for that channel)

For quantum mechanics $CH$ can be larger than zero, e.g. for a maximally
entangled state the largest value is obtained for $\theta _{1}=67^{o}.5$ , $%
\theta _{2}=45^{o}$, $\theta _{1}^{\prime }=22^{o}.5$ , $\theta _{2}^{\prime
}=0^{o}$ and correspond to a ratio 
\begin{equation}
R=[N(\theta _{1},\theta _{2})-N(\theta _{1},\theta _{2}^{\prime })+N(\theta
_{1}^{\prime },\theta _{2})+N(\theta _{1}^{\prime },\theta _{2}^{\prime
})]/[N(\theta _{1}^{\prime },\infty )+N(\infty ,\theta _{2})]  \label{eq:R}
\end{equation}
equal to 1.207.

For non-maximally entangled states the angles for which CH is maximal are
somehow different and the largest value smaller. The angles corresponding to
the maximum can be evaluated maximising Eq. \ref{eq:CH} with 

\bea
\left. \begin{array}{l}

  N[\theta _{1},\theta _{2}] =  [ \epsilon _1^{||} \epsilon _2^{||} (Sin[\theta _{1}]^{2}\cdot Sin[\theta_{2}]^{2}) + \\
  \epsilon _1^{\perp} \epsilon _2^{\perp} 
(Cos[\theta _{1}]^{2} \cdot Cos[\theta _{2}]^{2} )\\
  (\epsilon _1^{\perp} \epsilon _2^{||} Sin[\theta _{1}]^2\cdot Cos[\theta _{2}]^2 + \epsilon _1^{||} \epsilon _2^{\perp} 
Cos[\theta _{1}]^2 \cdot Sin [\theta _{2}]^2 )  \\
 + |f|^{2}\ast (\epsilon _1^{\perp} \epsilon _2^{\perp} (Sin[\theta _{1}]^{2}\cdot Sin[\theta_{2}]^{2}) +  \epsilon _1^{||} \epsilon _2^{||}
(Cos[\theta _{1}]^{2} Cos[\theta _{2}]^{2} ) +\\
(\epsilon _1^{||} \epsilon _2^{\perp} Sin[\theta _{1}]^2\cdot Cos[\theta _{2}]^{2} +\\
 \epsilon _1^{\perp} \epsilon _2^{||} 
Cos[\theta _{1}]^2 \cdot Sin [\theta _{2}]^2 )   \\
  +  (f+f^{\ast }) ((\epsilon _1^{||} \epsilon _2^{||} + \epsilon _1^{\perp} \epsilon _2^{\perp} - \epsilon _1^{||} \epsilon _2^{\perp} - 
\epsilon _1^{\perp} \epsilon _2^{||})
\cdot (Sin[\theta _{1}]\cdot Sin[\theta _{2}]\cdot
Cos[\theta _{1}]\cdot Cos[\theta _{2}]) ]  /(1+|f|^{2}) \,
\end{array}\right. \, .   
\eea
where (for the case of non-ideal polariser) $\epsilon _i^{||}$ and $\epsilon _i^{\perp}$ 
correspond to the transmission when the polariser (on the branch $i$)  axis is aligned or normal the polarisation axis respectively.

In our case we have a state with $f \simeq 0.4$ which corresponds,
for $\theta_1 =72^o.24$ , $\theta_2=45^o$, $\theta_1 ^{\prime}= 17^o.76$ and 
$\theta_2 ^{\prime}= 0^o$, to $R=1.16$.

Our experimental result is $CH = 512 \pm 135$ coincidences per second, 
which is almost four standard deviations different from zero and compatible with the theoretical value predicted by quantum mechanics.
In terms of the ratio (\ref{eq:R}), our result is $1.082 \pm 0.031$.

It must be noticed that results which are by far of many more standard
deviations above zero have been obtained. Nevertheless, to our knowledge,
this is the first measurement of the Clauser-Horne sum (or other Bell inequalities) using a non-maximally entangled state and thus represents and interesting result as a first step in the direction of eliminating the detection
loophole. This is also the first measurement of a polarisation two photon entanglemend
state, where the two photons have different wavelengths.
Further developments in this sense are the purpose of this collaboration.

\bigskip 

\vskip1cm \noindent {\bf Acknowledgements} \vskip0.3cm We would like to
acknowledge support of Italian Space Agency under contract LONO 500172.

\newpage
{\bf Figure caption}

Sketch of the source of polarisation entangled photons. CR1 and CR2 are two $LiIO_3$ crystals cut at the phase-matching angle of $51^o$. L1 and L2 are two identical piano-convex lenses with a hole of 4 mm in the centre. P is a 5 x 5 x 5 mm quartz plate for birefringence compensation and $\Lambda / 2$ is a first order half wave-length plate at 351 nm. 
U.V. identifies the pumping radiation at 351 nm. The infrared beam (I.R.) at 789 nm is generated by a diode laser and is used for system alignment only. The parametric amplifier scheme, described in the text, is shown as well.
The dashed line identifies the idler radiation at 633 nm. A second half-wave plate on the I.R. beam (not shown in the figure) allows amplified idler emission from the second crystals too. The figure is not in scale.

\end{document}